\documentclass{aa}
\usepackage{graphics,epsf,epsfig,graphicx}
\usepackage{colordvi}
\begin{document}
\title{Cold molecular gas in cooling flow clusters of galaxies} 
\author{P. Salom\'e
\inst{1} and F. Combes \inst{1}} \offprints {Philippe Salom\'e
(philippe.salome@obspm.fr)} \institute{Observatoire de Paris, LERMA,
61 Av. de l'Observatoire, F-75014 Paris, France} \date{Received ;
accepted} \authorrunning{Salom\'e \& Combes} \titlerunning{Search for
CO in central cluster galaxies}  \abstract{The results of a CO line
survey in central cluster galaxies with cooling flows are
presented. Cold molecular gas is detected with the IRAM 30m telescope,
through CO(1-0) and CO(2-1) emission lines in 6-10 among 32
galaxies. The corresponding gas masses are between 3.10$^{8}$ and
4.10$^{10}$ M$_\odot$. These results are in agreement with recent CO
detections by \cite{Edg01}. A strong correlation between the CO
emission and the H$\alpha$ luminosity is also confirmed. Cold gas
exists in the center of cooling flow clusters and these detections may
be interpreted as an evidence of the long searched very cold residual
of the hot cooling gas.}   
\maketitle
\section{Introduction}  Studies of
X-ray emission of hot intra-cluster medium (ICM) have pointed out the
high density of this gas in the central regions of many clusters. The
derived timescales for radiative cooling in the center is much smaller
than the Hubble time, and the ICM is predicted to condense and flow
towards the cluster center (see \cite{Fab94} for a review). The X-ray
spectra show evidence of cooler gas in the center, through central
drops of temperature.  But the fate of the cooled gas still remains
uncertain. The duration of the cooling flows is thought to be a
significant fraction of the cluster life-time, since cooling flows are
quite frequent in clusters. Estimated cooling rates of the order of
100 M$_\odot$/yr and up to 1000 M$_\odot$/yr implied that enormous
quantities of material should have accumulated (10$^{11}$ to 10$^{12}$
M$_\odot$ in a fraction of a Hubble time). But no resulting cold gas
has been detected in molecular form until recently.  Many efforts have
been expended to detect this gas in emission or absorption, either in
HI (\cite{Bur81}; \cite{Val82}; \cite{Sho83}; \cite{Mcn90};
\cite{Dwa95}) or in the CO molecule, see \cite{Grab90}; \cite{Mcn94};
\cite{Ant94}; \cite{Bra94}; \cite{Ode94}. The intracluster medium is
enriched in heavy elements with a metallicity of up to 0.3 solar
making possible the formation of CO molecules. The first detection of
CO emission has been made in Perseus A by \cite{Laz89}, but the
corresponding H$_2$ is not strongly identified as coming from the
cooling flow rather than from the galaxy itself. Recently,
\cite{Edg01} reported to have found CO line emission in the central
galaxy of sixteen extreme cooling flow clusters. Starbursts that may
appear as a consequence of the gas condensation must produce a lot of
young and hot stars. But the observed stellar luminosities are not
bright enough to account for the high mass deposition rates of cooling
flows. Although Chandra and XMM-Newton observations lead to reduced
rates, the cold molecular gas masses observed in some cluster cores
remain a small fraction of the gas cooled along the flow. The ICM is
probably multi-phase (e.g. \cite{Fer94}). A significant fraction of
gas might be so cold (\cite{Pfe94}) that it could correspond to the
high concentration of dark matter in clusters deduced from X-ray data
and gravitational arcs (\cite{Dur94}; \cite{Wu93}). Recently
\cite{Lie96}, \cite{Lie99} and \cite{Mit98} have detected large
quantities of gas at intermediate temperature of $5\cdot 10^5\,\rm K$
in 5 clusters with the EUVE satellite (Extreme Ultraviolet
Explorer). Since this phase is quite transient, the mass flow implied
would be much larger than that of the cooling flow itself. Other
processes must be at work, such as heating by shocks, or mixing layer
mechanisms at the interface between a cold and hot phase
(\cite{Bon01}). Also the detection of the near-infrared quadrupolar
emission line H$_2$(1-0)S(1) in central cluster galaxies with cooling
flows (and their non-detection in similar control galaxies without
cooling flows, e.g. \cite{Fal98}) support the presence of molecular
gas at temperature of 2000K (\cite{Jaf97}, Edge et al. 2002,
\cite{Wil02}).\\ In this paper, we present our search for CO lines in
32 galaxies in the center of clusters, carried out in June and August
2001 with the IRAM 30m telescope. We have found 6 clear detections and
4 hints of CO lines. In the next section we describe the instrumental
conditions of our observations and the data reduction. We then present
results and cold gas mass evaluations in section 3. In section 4 and 5
we discuss the possible significations of such large gas quantities
when they are present and compare these measurements with other
wavelength
observations.  
\small 
\begin{table*}[htbp]
\centering 
\begin{tabular}{lccccccc}
\hline
\hline 
\multicolumn{1}{l}{Source }& \multicolumn{1}{c}{Redshift} &
\multicolumn{1}{c}{RA (2000)} & \multicolumn{1}{c}{Decl (2000)} &
\multicolumn{1}{c}{Central Freq.} &\multicolumn{1}{c}{Exposure}&
\multicolumn{1}{c}{Central Freq.  } &\multicolumn{1}{c}{Exposure} \\
\multicolumn{1}{l}{}& \multicolumn{1}{c}{} & \multicolumn{1}{c}{} &
\multicolumn{1}{c}{} & \multicolumn{1}{c}{in C0(1-0)}
&\multicolumn{1}{c}{time (min)}& \multicolumn{1}{c}{in CO(2-1)}
&\multicolumn{1}{c}{time (min)} \\ 
\hline
A85$^{*}$ & 0.05567 & 00 41
50.4 & -09 18 11 & 109.19 GHz & 152 & - &- \\ Z235 & 0.08300 & 00 43
52.1 & +24 24 21 & 106.43 GHz & 128 & 212.87 GHz & 128 \\ A262$^{*}$ &
0.01620 & 01 52 46.5 & +36 09 07 & 113.43 GHz & 160 & - &- \\
A291$^{*}$ & 0.19590 & 02 01 43.1 & -02 11 48 & 96.38 GHz & 304 & - &
- \\ A496 & 0.03281 & 04 33 37.8 & -13 15 43 & 111.60 GHz & 248 &
223.21 GHz & 256 \\ RXJ0439+05 & 0.20800 & 04 39 02.2 & +05 20 44 &
95.42 GHz & 300 & - & - \\ PKS0745-191 & 0.10280 & 07 47 31.3 & -19 17
40 & 104.52 GHz & 304 & 209.04 GHz & 376 \\ A644$^{*}$ & 0.07040 & 08
17 25.5 & -07 30 44 & 107.69 GHz & 144 & 215.37 GHz & 88 \\ RXJ0821+07
& 0.11000 & 08 21 02.4 & +07 51 47 & 103.84 GHz & 96 & 207.69 GHz & 96
\\ A646 & 0.12680 & 08 22 09.6 & +47 05 53 & 102.30 GHz & 304 & 204.59
GHz & 312 \\ A780 & 0.05384 & 09 18 05.7 & -12 05 44 & 109.38 GHz &
268 & 218.76 GHz & 352 \\ A1068$^{*}$ & 0.13860 & 10 40 44.5 & +39 57
11.1 & 101.23 GHz & 180 & - &- \\ A978 & 0.05425 & 10 20 26.5 & -06 31
36 & 109.34 GHz & 96 & 218.67 GHz & 96 \\ A1668 & 0.06368 & 13 03 46.6
& +19 16 18 & 108.37 GHz & 80 & 216.73 GHz & 80 \\ 
A1795 & 0.06326 & 13 48 52.4 & +26 35 34 & 108.41 GHz & 68 & 216.82 GHz & 129
\\ A2029 & 0.07795 & 15 10 56.1 & +05 44 41 & 106.93 GHz & 112 &
213.86 GHz & 168 \\ MKW3s & 0.04531 & 15 21 51.9 & +07 42 32 & 110.27
GHz & 144 & 220.54 GHz & 144 \\ A2146 & 0.23370 & 15 56 13.8 & +66 20
55 & 93.43 GHz & 272 &- & - \\ A2142$^{*}$ & 0.09037 & 15 58 20.0 &
+27 14 02 & 105.71 GHz & 128 &- &- \\ A2147 & 0.03532 & 16 02 17.0 &
+15 58 28 & 111.33 GHz & 152 & 222.67 GHz & 152 \\ A2151 & 0.03533 &
16 04 35.8 & +17 43 18 & 111.33 GHz & 152 & 222.67 GHz & 152 \\ A2199
& 0.03035 & 16 28 38.5 & +39 33 06 & 111.87 GHz & 128 & 223.74 GHz &
120 \\ 
Z8193$^{*}$ & 0.18290 & 17 17 19.2 & +42 27 00 & 97.44
GHz & 304 &- &- \\ A2261$^{*}$ & 0.2240 & 17 22 27.1 & +32 07 58 &
94.17 GHz & 256 &- &- \\ A2319$^{*}$ & 0.05459 & 19 21 10.0 & +43 56
44 & 109.30 GHz & 136 & 218.60 GHz & 88 \\ CygA & 0.05607 & 19 59 28.3
& +40 44 02 & 109.15 GHz & 228 & 218.29 GHz & 232 \\ A2462 & 0.07437 &
22 39 11.4 & -17 20 28 & 107.29 GHz & 144 & 214.58 GHz & 144 \\ A2597
& 0.08520 & 23 25 19.8 & -12 07 26 & 106.22 GHz & 144 & 212.43 GHz &
144 \\ A2626 & 0.05490 & 23 26 30.6 & +21 08 50 & 109.27 GHz & 136 &
218.54 GHz & 140 \\ A2634 & 0.03022 & 23 38 29.5 & +27 01 56 & 111.89
GHz & 144 & 223.77 GHz & 144 \\ A2657 & 0.04023 & 23 44 57.4 & +09 11
34 & 110.81 GHz & 208 & 221.62 GHz & 208 \\ A2665$^{*}$ & 0.05610 & 23
50 50.6 & +06 09 00 & 109.14 GHz & 192 & 218.29 GHz & 192 \\
\hline 
\end{tabular}
\caption{This table presents the sample of
cooling flow clusters of galaxies observed with the IRAM 30m
telescope. Central frequencies of the CO(1-0) and CO(2-1) lines
observed, as the exposure time are indicated for each galaxy. Sources
observed in the second run (August) are with indicated by a star. The
others were observed during the first run (July). Several sources were
not observed in CO(2-1) since the redshifted J=2-1 transition lines
were out of the 30m telescope receiver's
band.} 
\label{prop1} 
\end{table*}  

\begin{table*}[htbp] 
\centering \begin{tabular}{lccccccc} 
\hline
\hline 
\multicolumn{1}{c}{Source } & \multicolumn{1}{c}{Line}
& \multicolumn{1}{c}{Peak} & \multicolumn{1}{c}{Rms}
& \multicolumn{1}{c}{Line} &\multicolumn{1}{c}{Line}
& \multicolumn{1}{c}{Line} & \multicolumn{1}{c}{I$_{CO}$}
\\ \multicolumn{1}{c}{} & \multicolumn{1}{c}{} &
\multicolumn{1}{c}{mK} & \multicolumn{1}{c}{mK} &
\multicolumn{1}{c}{detection } & \multicolumn{1}{c}{position (km/s)} &
\multicolumn{1}{c}{width (km/s)} & \multicolumn{1}{c}{K.km/s} \\
\hline A85$^{*}$ & CO(1-0) & -& 0.9 & no & - & 300$^\star$ &$\le$0.35
\\ Z235 & CO(1-0) & 1.9$\pm$0.2& 0.8 & hint & -258$\pm$48 &
318$\pm$103& 0.66$\pm$0.19 \\ & CO(2-1) & - & 1.7 & no & - &
300$^\star$ & $\le$0.65 \\ A262$^{*}$ & CO(1-0) & 2.9$\pm$0.5 & 0.8 &
yes & 31$\pm$24 & 346$\pm$47& 1.05$\pm$0.14 \\ A291$^{*}$ & CO(1-0) &
- & 0.9 & no & - & 300$^\star$ & $\le$0.35 \\ A496 & CO(1-0) &
1.5$\pm$0.5 & 0.7 & hint & 382$\pm$39 & 311$\pm$73 & 0.49$\pm$0.12 \\
& CO(2-1) & 3.0$\pm$0.7 & 1.1 & hint & 114$\pm$29 & 249$\pm$52 &
0.80$\pm$0.18 \\ RXJ0439+05 & CO(1-0) & - & 0.72 & no & - &
300$^\star$ & $\le$0.28 \\ PKS0745-191& CO(1-0) & 2.0$\pm$0.3 & 0.6 &
yes & 18$\pm$29 & 221$\pm$58 & 0.47$\pm$0.11 \\ & CO(2-1)
& 11.2$\pm$0.8 & 2.1 & yes & -45$\pm$15 & 215$\pm$41 & 2.57$\pm$0.38
\\ A644$^{*}$ & CO(1-0) & 1.1$\pm$0.4 & 0.7 & no & -10$\pm$55
& 260$\pm$77 & $\le$0.27 \\ & CO(2-1) & - & 3.1 & no & - & 300$^\star$
& $\le$1.12 \\ RXJ0821+07 & CO(1-0) & 8.9$\pm$0.6 & 1.1 & yes &
-2$\pm$ 8 & 135$\pm$21 & 1.28$\pm$0.16 \\ & CO(2-1) & 9.9$\pm$1.5 &
2.9 & yes & 5$\pm$34 & 270$\pm$81 & 2.86$\pm$0.60 \\ A646 & CO(1-0)
& 1.5$\pm$0.1 & 0.5 & yes & 105$\pm$39 & 376$\pm$121& 0.62$\pm$0.15 \\
& CO(2-1) & 1.9$\pm$1.0 & 1.6 & no & 46$\pm$54 & 346$\pm$123&
$\le$0.62 \\ A780 & CO(1-0) & 1.8$\pm$0.5 & 0.6 & yes & 219$\pm$31 &
439$\pm$68& 0.80$\pm$0.11 \\ & CO(2-1) & - & 1.8 & no & - &
300$^\star$ & $\le$0.69 \\ A978 & CO(1-0) & 1.8$\pm$0.7 & 1.0 & no &
-128$\pm$49 & 221$\pm$87 & $\le$0.39 \\ & CO(2-1) & - & 2.1 & no & - &
300$^\star$ & $\le$0.80 \\ A1068$^{*}$ & CO(1-0) & 10.1$\pm$0.3 & 0.5
& yes & -45$\pm$4 & 249$\pm$10 & 2.66$\pm$0.1 \\ A1668 & CO(1-0) & - &
1.2 & no & - & 300$^\star$ & $\le$0.46 \\ & CO(2-1) & 2.4$\pm$1.6& 1.7
& no & 9$\pm$64 & 281$\pm$125& $\le$0.65 \\ 
A1795 & CO(1-0) &
3.4$\pm$0.7 & 0.9 & yes & -190$\pm$28 & 405$\pm$56 &  1.47$\pm$0.19 \\
& CO(2-1) & 6.1$\pm$1.0 & 1.8 & yes &  -128$\pm$31 & 500$\pm$65 &
3.26$\pm$0.39 \\ 
\hline 
\end{tabular} 
\caption{Summary of
observational data for the two runs. Lines characteristics are
presented, spectra are shown on Fig. \ref{spectres}. I$_{CO}$ (column
8) were evaluated from a gaussian fit of the CO(1-0) and CO(2-1) lines
for detections and hints of detections. I$_{CO}$ upper limits were
evaluated by equation \ref{Ico} for non-detections. A detection is
asserted in one transition line when the peak of the gaussian fit is
above three times the rms (in 55 km/s channels) and a hint of
detection when it is between two and three times the rms. To claim a
cold molecular gas detection, we require a detection in both
transition lines. A possible detection was claimed when was present a
detection in one transition line or a hint of detection in both lines,
unless the line appear clearly in one transition only (Abell 262,
Abell 1068, Zw8193).}
\label{prop2} 
\end{table*}  
\begin{table*} 
\centering 
\begin{tabular}{lccccccc} 
\hline
\hline
\multicolumn{1}{c}{Source } & \multicolumn{1}{c}{Line}
& \multicolumn{1}{c}{Peak} & \multicolumn{1}{c}{Rms}
& \multicolumn{1}{c}{Line} &\multicolumn{1}{c}{Line}
& \multicolumn{1}{c}{Line} & \multicolumn{1}{c}{I$_{CO}$}
\\ \multicolumn{1}{c}{} & \multicolumn{1}{c}{} &
\multicolumn{1}{c}{mK} & \multicolumn{1}{c}{mK} &
\multicolumn{1}{c}{detection } & \multicolumn{1}{c}{position (km/s)} &
\multicolumn{1}{c}{width (km/s)} & \multicolumn{1}{c}{K.km/s} \\
\hline  
A2029 & CO(1-0) & 1.1$\pm$0.5 & 0.9 & no & -68$\pm$87 &
445$\pm$168 & $\le$0.35 \\ & CO(2-1) & - & 1.6 & no & - & 300$^\star$
& $\le$0.62 \\ MKW3s & CO(1-0) & 1.3$\pm$0.4 & 0.8 & no & -133$\pm$60
& 347$\pm$100 & $\le$0.31 \\ & CO(2-1) & - & 1.0 & no & - &
300$^\star$ & $\le$0.39 \\ A2146 & CO(1-0) & - & 0.5 & no & - &
300$^\star$ & $\le$0.19 \\ A2142$^{*}$ & CO(1-0) & - & 0.7 & no & - &
300$^\star$ & $\le$0.27 \\ A2147 & CO(1-0) & - & 1.1 & no & - &
300$^\star$ & $\le$0.42 \\ & CO(2-1) & - & 1.4 & no & - & 300$^\star$
& $\le$0.54 \\ A2151 & CO(1-0) & 2.0$\pm$0.3 & 0.8 & hint & 341$\pm$47
& 133$\pm$95 & 0.29$\pm$0.14 \\ & CO(2-1) & - & 1.5 & no & - &
300$^\star$ & $\le$0.58 \\ A2199 & CO(1-0) & 1.6$\pm$0.4 & 0.9 & no &
109$\pm$55 & 395$\pm$123 & $\le$0.35 \\ & CO(2-1) & 3.1$\pm$0.6 & 1.2
& hint & 265$\pm$28 & 204$\pm$61 & 0.67$\pm$0.18 \\ 
Z8193$^{*}$ & CO(1-0) & 2.1$\pm$0.4 & 0.4 & yes &
-14$\pm$20 & 242$\pm$65 & 0.55$\pm$0.1 \\ A2261$^{*}$ & CO(1-0) & - &
0.8 & no & - & 300$^\star$ & $\le$0.31 \\ A2319$^{*}$ & CO(1-0) & - &
0.7 & no & - & 300$^\star$ & $\le$0.27 \\ & CO(2-1) & - & 3.4 & no & -
& 300$^\star$ & $\le$1.3 \\ CygA & CO(1-0) & 3.6$\pm$0.3 & 1.4 & hint
& 203$\pm$24 & 153$\pm$46 & 0.59$\pm$0.18 \\ & CO(2-1) & 2.1$\pm$0.5
& 1.4 & no & -83$\pm$65 & 336$\pm$207 & $\le$0.54 \\ A2462 & CO(1-0) &
- & 0.9 & no & - & 300$^\star$ & $\le$0.35 \\ & CO(2-1) & - & 2.7 & no
& - & 300$^\star$ & $\le$1.0 \\ A2597 & CO(1-0) & 1.1$\pm$0.3 & 0.8 &
no & -10$\pm$55 & 260$\pm$77 & 0.29$\pm$0.11 \\ & CO(2-1) & - & 2.3 &
no & - & 300$^\star$ & $\le$0.89 \\ A2626 & CO(1-0) & 1.7$\pm$0.2 &
0.8 & hint & -161$\pm$42 & 293$\pm$103 & 0.52$\pm$0.15 \\ & CO(2-1) &
- & 1.4 & no & - & 300$^\star$ & $\le$0.54 \\ A2634 & CO(1-0)
& 2.9$\pm$0.4 & 1.1 & hint & 488$\pm$62 & 459$\pm$137 & 1.44$\pm$0.38
\\ & CO(2-1) & - & 2.2 & no & - & 300$^\star$ & $\le$0.85 \\ A2657
& CO(1-0) & 3.0$\pm$0.03& 0.8 & yes & 147$\pm$13 & 69$\pm$50
& 0.22$\pm$0.080 \\ & CO(2-1) & - & 0.8 & no & - & 300$^\star$
& $\le$0.31 \\ A2665$^{*}$ & CO(1-0) & 1.5$\pm$0.2 & 0.8 & no
& -273$\pm$40 & 193$\pm$82 & $\le$0.31 \\ & CO(2-1) & - & 1.6 & no & -
& 300$^\star$ & $\le$0.62 \\ 
\hline 
\end{tabular} 
\caption{Summary of
observational data for the two runs. Lines characteristics (Table
\ref{prop2}
continuation).}
\label{prop3} 
\end{table*} 
\normalsize  
\section{Observations
and data reduction}  The sample of sources was selected according to
several criteria. First, we wanted to observe galaxies with important
cooling flows, so we chose high deposition rates galaxies with
$\dot{M}$ around or greater than 100 M$\odot$/yr, see \cite{Per98},
\cite{whi97}, though these rates are certainly overestimated. Three
non-cooling flow clusters (Abell 1668, Abell 1704 and Abell 2256) have
also been observed and not detected in CO with the 30m telescope. It
is possible that the large gas flow produces massive stars ionizing
the gas. The gas might also be cooling in ionizing shocks
(optically luminous). Thus, the presence of large amounts of cooled
gas could be accompanied by H$\alpha$ emission as suggested
in \cite{Edg01}. Sources were then selected according to
their H$_{\alpha}$ luminosity when available (high luminosity of
about $10^{42}$ erg.$s^{-1}$ from \cite{Cra99}, \cite{Owe95}). The
sample contains only relatively low-redshift cD galaxies ($z < 0.25$),
for the sake of sensitivity. We gather data at other wavelengths, such
as the far infra-red, when available, to be able to compare gas and
dust emission. All observing parameters are summarized in
Table \ref{prop1}. Observations were achieved with the IRAM
30m millimeter-wave telescope at Pico Veleta, Spain in June and
August 2001 in good weather conditions. We used four
receivers simultaneously, centered two on the CO(1-0) and two on the
CO(2-1) lines at 115 GHz and 230 GHz. The beam of the telescope at
these two frequencies is 22" and 13" respectively.  Two backends were
provided by the autocorrelator, with a 1.25 MHz resolution on a 600
MHz band width. The two other backends were the two 512MHz wide
1MHz filter-banks. These yield a total band of $\sim$ 1300km/s at
2.6mm and $\sim$ 650km/s at 1.3mm. In addition, we used the 4 MHz
resolution filter-bank, providing a 1 GHz band width, important for
the 1.3mm receivers (since it corresponds to 1300km/s bandwidth
also). Given the uncertainty in the central velocity of the CO line
(some optically measured velocities being systematically displaced
with respect to the galaxy systemic velocity), the expected width of a
cD galaxy, and the required baseline to eliminate sinusoidal
fluctuations, this wide band is necessary. The signals are expressed
in main beam temperatures, since the sources are not expected to be
extended and homogeneous. The main-beam efficiency of the 30m is
:  \begin{equation} \eta_{\rm mb}=T_{\rm A}^*/T_{\rm mb} \quad {\rm
and} \quad \eta_{\rm mb}=B_{eff}/F_{eff} \end{equation}  with
$\eta_{\rm mb}$= 0.75/0.95 at 115 GHz and 0.52/0.91 at 230 GHz (cf
IRAM-30m site http://www.iram.es/). The data were reduced with
the CLASS package; spiky channels and bad scans were excised.
After averaging all the raw spectra for each line of each source,
linear baselines were subtracted and the spectra were
Hanning smoothed. Assuming that a good sampling of the line requires
at least five points, and assuming a typical 300 km/s line width, the
data were smoothed to 55 km/s of spectral resolution, at 1.3 and
2.6mm, to gain more than a factor 3 in signal to noise. The CO
emission lines were fitted with gaussian profiles through the CLASS
package. For non detections, CO intensity I$_{\rm CO}$ upper limits at
3$\sigma$ were evaluated as in \cite{Mcn94} by
:  \begin{equation} {\rm I}_{\rm CO}=3 \times rms \times W_{line}
\times (\frac{W_{sm}}{W_{line}})^{1/2}
\label{Ico} \end{equation}  where rms is the noise level, computed in
the channel width $W_{sm}$ (55 km/s) and $W_{line}$ is the expected
line bandwidth, typically 300 km/s. The typical integration time was
two hours for all galaxies. CO intensities were obtained by averaging
the data of at least two different days, to prevent systematic effects
(baseline ripples...). The detection criterion is the CO emission line
maximum is at least three times the rms, in 55km/s channels.  When the
signal to noise ratio was between 2 and 3 rms we concluded to a hint
of detection. Below 2 rms, upper limits were computed for the
CO intensity. Results are presented in Table \ref{prop2} and
Table \ref{prop3}. After this first analysis at each wavelength, we
compare I$_{CO(1-0)}$ and I$_{CO(2-1)}$. A cold gas detection is
claimed when CO is detected in the CO(1-0) and in the CO(2-1)
lines simultaneously. Possible detections are defined by a detection
in one line or by a hint of detection in both lines. Otherwise, we
concluded to a non detection. For Abell 262, Abell 1068 and Zw8193,
without CO(2-1) data, a detection is claimed because of a clear line
detection in CO(1-0). With these criteria, we claim 6 detections, 4
possible detections, and 22 non detections. Detections made by
\cite{Edg01} for Abell 1068 and RXJ0821+07 are confirmed, with a good
agreement of the derived hydrogen molecular gas masses. CO(1-0)
emission line is also confirmed in Abell 262, but molecular gas mass
deduced here is twice lower than in \cite{Edg01}. That comes from the
fact we identify a line with a smaller width. 
Three new values of M$_{gas}$ are found :
Abell 646 for which cold gas mass is in agreement with \cite{Edg01}
upper limit, Abell 1795 for which a large line width is found and
consequently a molecular hydrogen mass higher than the upper limit
deduced in \cite{Edg01} and PKS0745-19 that was observed at a wrong
frequency by \cite{Edg01}. 

CO intensities always correspond to areas
deduced from gaussian fits for detections and possible detections when
a gaussian fit was possible, but for the 22 no-detections, the
I$_{co}$ are evaluated with Formula (2). Table \ref{results} show gas
mass estimates as well as X-ray, optical, IR and radio data when
available.  \section{Results}   \subsection{H$_2$ mass evaluation from
CO observations}  
\begin{figure}[h] 
\centerline{\epsfig{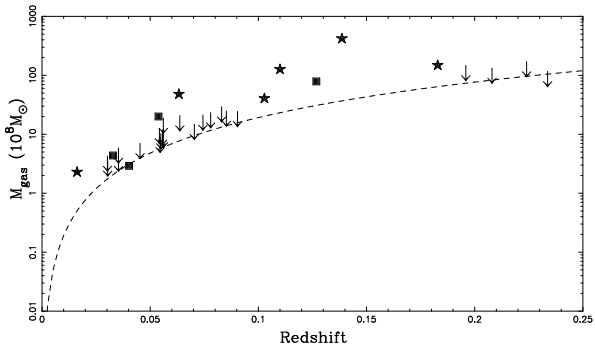}} 
\caption{M$_{gas}$in 10$^8$M$_{\odot}$ deduced from CO observations, versus the galaxy
redshift. Filled stars are detections, filled square hints of
detection and arrows are upper limits. The dashed line represents the
molecular mass limit than can be deduced from CO observations with the
IRAM 30m telescope, in 2h integration
time.} \label{red} 
\end{figure}

Since cold H$_2$ is a symmetric molecule, the best
tracer of cold molecular gas is the CO lines, from the most abundant
molecule after H$_2$: CO/H$_2 \sim 6.10^{-5}$.  From standard (and
empirical) calibrations, it is possible to deduce the interstellar
H$_2$ content from the integrated CO intensity I$_{\rm CO}$ (K.km/s)
:  \begin{equation} {\rm I}_{\rm CO}= \int T_{\rm mb}(CO)\,
dV \end{equation}  where $T_{\rm mb}(CO)$ is the main beam antenna
temperature, obtained for the CO(1-0) line (cf Sect. 2). Although the
typical molecular cloud is optically thick in the first CO lines of
the J ladder, the proportionality factor between the column density of
molecular gas and integrated intensity is justified, since the
observed signal is the emission sum of many clouds in the beam, and
these clouds have a small filling factor, when spatial and velocity
volume is considered. We adopt here the conversion factor commonly
used for N(H$_2$) in molecule/cm$^2$ unit :  \begin{equation} N(H_2) =
2.3 \, 10^{20} \, {\rm I}_{\rm CO} \end{equation}  From this equation
the mass of molecular hydrogen, contained in one beam, in M$_{\odot}$
is:  \begin{equation} {\rm M(H}_2) = 2.95 \, 10^{-19} \, {\rm I}_{\rm
CO}\, \theta^2 \, D^2 \, \frac{N(H_2)}{{\rm I}_{\rm
CO}} \end{equation}  where I$_{\rm CO}$ is the integrated intensity in
K.km/s, $\theta$ is the beamsize of the telescope in arcseconds, and D
is the distance of the galaxy in Mpc, determined with a Hubble
constant $H_0$=70 km/s/Mpc. This converting factor is in agreement
with the one used in \cite{Edg01} and in previous CO observations of
cluster cores but gas mass estimations are slightly lower here because
we took $H_0$=70 km/s/Mpc. It is important to notice the
I$_{CO}$/N(H$_2$) conversion factor has been first calibrated in the
solar neighborhood. To use this value implies intra-cluster medium is
assumed to behave like the Galactic interstellar matter near the sun,
in particular with the solar metallicity. But the intracluster medium
has subsolar metallicity. So this conversion factor is likely to
underestimate the mass of molecular hydrogen.  A standard factor 1.36
taking into account He contribution is also used in the gas mass
estimation : M$_{gas}$=1.36M(H$_2$). The derived M$_{gas}$ values,
displayed in Table \ref{results}, are between $\sim$ 10$^8$ and $\sim$
10$^{10}$ M$_\odot$. These masses, together with upper limits, are
plotted on Fig. \ref{red} for all sources, as a function of
redshift. The curve represents the M$_{gas}$detection limit of the
IRAM 30m telescope for the CO(1-0) line, assuming a typical 300 km/s
linewith and a temperature detection limit of 0.5 mK, in agreement
with our noise level.  
\subsection{CO
detections}  \textbf{Abell 262} has also been detected by
\cite{Edg01}. Current values are compatible with these measurements
and confirm this detection even if CO(2-1) data were not good enough
in the second run of August 2001 to detect a line. This galaxy
contains a central radio source according to \cite{Per98} and show low
luminosity optical lines, see \cite{Cra99}.\\ \textbf{PKS 0745-191} is
a 0.1028 redshifted galaxy already observed in CO by \cite{Edg01}. In
present observations, a CO(1-0) line is detected three times above the
rms. A simultaneous detection in the CO(2-1) band with a signal to
noise better then five seems to confirm the presence of molecular
gas. This galaxy is supposed to contain a large cooling flow with
mass deposition rates around 1000 M$_\odot$/yr according to
\cite{Per98}, \cite{All00}. Strong optical emission lines have also
been detected in PKS 0745-191. This galaxy is the site of an important
excitation mechanism. Besides it is a powerful radio source with an
amorphous and filamentary morphology, see \cite{Bau91}.  Recently,
\cite{Don00} have mapped kpc-size filaments in vibrationally-excited
H$_2$ in the cores of galaxies centers of cooling flows, like PKS
0745-191, with high spatial resolution.  They have also found dust
lanes which are optically thick to 1.6$\mu$m emission. These dust
lanes are confined to the central few kpcs.\\ The cD galaxy
\textbf{RXJ0821+07} has been detected in CO by \cite{Edg01}. This
relatively easy detection is confirmed here in both wavelength and CO
intensities are compatible with previous ones. Optical images taken
with the AAT and Hubble Space Telescope by \cite{Bay02} show that the
central galaxy is embedded in a luminous and extended line-emitting
nebula coincident with a bright excess of X-ray emission imaged by
Chandra. \\ \textbf{Abell 1068} detected by \cite{Edg01} is confirmed
here.  The H$_2$ mass deduced is the highest we found among
detections.  This galaxy show strong optical lines and large dust mass
in comparison with other detections. It is also a powerful IRAS source
with a 650 mJy flux at 60 $\mu$m.\\ \textbf{Abell 1795} has been
observed by \cite{Bra94} who did not detect CO line. \cite{Edg01}
found a marginal detection. This detection is confirmed here in the
two bands. But line widths deduced from gaussian fits are quite
different in the two wavelengths, so we cannot exclude the possibility
of very high velocity molecular clouds. This galaxy is known as a
radio source see \cite{Dav93}. An optical filament has been detected
in H$_\alpha$ by \cite{Cow83}. According to them, some of the
filaments observed in Abell 1795 seem to be concentrated and coming
from the galaxy whereas fainter extended filaments are surrounding the
galaxy. The question about their origin and their link with the
cooling flows is not clearly determined. Mapping cold gas will allow
to better understand the spatial structure of the cooling material and
to know if the CO is along the filaments or in the galaxy, Salom\'e \&
Combes (2003, in prep).\\ \textbf{Zwicky 8193} is a strong optical
line emitter. Only CO(1-0) was observed here. The molecular gas mass
deduced agree with the value derived in \cite{Edg01}. Zwicky 8193 is a
complex system and we refer to \cite{Edg01} and \cite{Edg02} for the
discussion.   \subsection{Hints of detection}  We consider 4 galaxies
of our sample to be possible CO emitters according to criteria defined
above. Nevertheless CO emission lines here are fainter than the
previous ones, with values reaching half a K.km/s.\\ \textbf{Abell
496} is a cD galaxy. A possible line is seen in the two bands, but
signal to noise ratio between 2 and 3 is not sufficient to claim a
detection. Much time has been dedicated to this radio source, see for
example \cite{Per98}, to deduce a small upper limit of H$_2$
mass. Faint optical H$_\alpha$ line have been observed in this galaxy
also emitting in X-ray, see \cite{Dav93}.  Nevertheless we deduce here
a new upper limit in molecular gas mass.\\ We also assert a possible
detection in \textbf{Abell 646}, even if no CO(2-1) line is seen,
because of the clear shape of the CO(1-0) line detected just above
three time the rms. Moreover, \cite{Edg01} asserted to have a marginal
detection of this galaxy.  So it would be interesting to confirm this
detection.\\ \textbf{Abell 780} (Hydra A) is a very powerful radio
source that had already been observed through millimetric wavelength,
see for example \cite{Ode94b}. This much studied source is here at the
limit of detection in CO(1-0) and not seen in CO(2-1). M$_{gas}$ upper
limit deduced is in agreement with the evaluation made by
\cite{Edg01}, but no clear detection can be claimed. \\ The cD
\textbf{Abell 2657} galaxy with optical emission lines was not
detected in CO(2-1). Faint possible CO(1-0) line is present and a new
upper limit in M$_{gas}$ is derived, but more observations are
required.\\   \subsection{Upper limits}  A large number of the
selected galaxies are not detected. It is possible, these cooling flow
clusters of galaxies contains cold gas with lines still too weak to be
detectable with the actual IRAM 30m telescope
sensitivity.  
\small 
\begin{table*}[htbp] 
\centering 
\begin{tabular}{llrrrrrrc} 
\hline
\hline 
\multicolumn{1}{l} {Source} & \multicolumn{1}{c} {M$(H_2)$} &
\multicolumn{1}{c} {Mdust} &\multicolumn{1}{c} {L(H$_{\alpha}$)}&
\multicolumn{1}{c} {$\dot{M}^{(1)}$}& \multicolumn{1}{c}
{r$_{cool}^{(1)}$}& \multicolumn{1}{c} {$\dot{M}^{(2)}$}&
\multicolumn{1}{c} {r$_{cool}^{(2)}$}& \multicolumn{1}{c}
{Flux(1.4GHz)}\\ \multicolumn{1}{l} {} & \multicolumn{1}{c}
{(M$_{\odot}$)} & \multicolumn{1}{c} {(M$_{\odot}$)}
&\multicolumn{1}{c} {(erg/s)}& \multicolumn{1}{c} {(M$_\odot$/yr)}&
\multicolumn{1}{c} {(kpc)}& \multicolumn{1}{c} {(M$_\odot$/yr)}&
\multicolumn{1}{c} {(kpc)}&\multicolumn{1}{c} {(mJy)}\\
\hline 
A262 & 2.3$\pm0.3$$\times$10$^8$ & 8.7$\times$10$^6$ &
6.0$\times$10$^{39}$ & 27& 104& 10& 67& 131 \\ pks0745-191 &
4.0$\pm0.9$$\times$10$^{9}$ & 3.5$\times$10$^7$ & 1.4$\times$10$^{42}$
& 1038 & 214&579 & 177&2370 \\ rxj0821+07 &
1.3$\pm0.2$$\times$10$^{10}$ & 4.2$\times$10$^7$
& 3.0$\times$10$^{41}$ & -& -& -& -& - \\ A1068
& 4.2$\pm0.2$$\times$10$^{10}$ & 1.4$\times$10$^8$
& 1.7$\times$10$^{42}$ & -&- &- &- & - \\ A1795
& 4.8$\pm0.6$$\times$10$^9$ & 6.7$\times$10$^7$ & 1.1$\times$10$^{41}$
& 381& 177& 321& 181& 930 \\ Z8193 & 1.5$\pm0.3$$\times$10$^{10}$
& 3.8$\times$10$^7$ & 1.5$\times$10$^{42}$ & -& -& -& -& - \\
\hline 
 A496 & 4.3$\pm$1.0$\times$10$^{8}$ & - &
3.4$\times$10$^{42}$ & 95& 110& 134&138 & - \\ A646 &
7.9$\pm0.2$$\times$10$^{9}$ & 2.5$\times$10$^7$ & 1.6$\times$10$^{41}$
& - &- &- &- & - \\ A780 & 2.0$\pm0.3$$\times$10$^9$ &
3.0$\times$10$^7$ & 1.6$\times$10$^{41}$ & 262&162 & 222&170 & 40800
\\ A2657 & 2.9$\pm0.1$$\times$10$^8$ & - & - & -&- & 44& 101&- \\
\hline 
A85 & $\le$ 8.8$\times$10$^{8}$ & - & - & 107 & 93& 108& 131&
58 \\ Z235 & $\le$ 2.5$\times$10$^{9}$ & - & - & -& -& -& -& - \\ A291
& $\le$ 1.1$\times$10$^{10}$& 84.9$\times$10$^7$ &
4.6$\times$10$^{41}$ & - &- & -& -& - \\ A644 & $\le$
1.1$\times$10$^{9}$ & - & - & 189& 141& 136& 111& - \\ rxj0439+05&
$\le$ 9.8$\times$10$^{9}$ & 5.1$\times$10$^7$ & 1.1$\times$10$^{42} $
& -& -& -& -& - \\ A978 & $\le$ 9.3$\times$10$^{8}$ & - &
2.0$\times$10$^{40}$ & - &- &- &- & -\\ A1668 & $\le$
1.5$\times$10$^{9}$ & - & 1.2 $\times$10$^{41}$ & -& -& -& -& -
\\ 
A2029
& $\le$ 1.7$\times$10$^{9}$ & - & 8.0$\times$10$^{39}$ & 555 & 186&
431& 192&550 \\ MKW3s & $\le$ 5.2$\times$10$^{8}$ & -
& 5.0$\times$10$^{39}$ & 175& 171& 132&158 & - \\ A2146 &
$\le$ 8.6$\times$10$^{9}$ & 8.8$\times$10$^7$ & 1.4$\times$10$^{42}$ &
-& -& -& -& - \\ A2142 & $\le$ 1.8$\times$10$^{9}$ & - & - & 350 &
150&369 &172 & - \\ A2147 & $\le$ 4.3$\times$10$^{8}$ & -
& 7.0$\times$10$^{39}$ & -& -& -& -& - \\ A2151 &
$\le$ 3.1$\times$10$^{8}$ & - & 5.8$\times$10$^{40}$ & -& -& 166&
146&- \\ A2199 & $\le$ 2.6$\times$10$^{8}$ & - & 3.5$\times$10$^{40}$
& 154 & 143& 97&124 & 3700 \\
A2261 & $\le$
1.3$\times$10$^{10}$& - & - & -& 20& 53& -& - \\ A2319 & $\le$
6.6$\times$10$^{8}$ & - & 1.0$\times$10$^{41}$ & -& -& -& -& - \\ CygA
& $\le$ 1.4$\times$10$^{9}$ & - & 6.5$\times$10$^{42}$ & 244& 135&
242& 167&- \\ A2462 & $\le$ 1.6$\times$10$^{9}$ & -
& 5.8$\times$10$^{40}$ & -& -& -& -&- \\ A2597 & $\le$
1.8$\times$10$^9$ & 8.4$\times$10$^7$ & 5.2$\times$10$^{41}$ & 271 &
152& -& -& 1880 \\ A2626 & $\le$ 7.6$\times$10$^{8}$ & - &
3.3$\times$10$^{40}$ & -& -& 53&114 &- \\ A2634 & $\le$
3.2$\times$10$^{8}$ & - & 3.7$\times$10$^{40}$ & -& -& -& -& 7657 \\
A2665 & $\le$ 7.9$\times$10$^{8}$ & - & 6.0$\times$10$^{39}$ & -& -&
-& -& - \\ 
\hline 
\end{tabular} 
\caption{Derived parameters of the
observed sources. Optical line luminosities are from
\cite{Cra99}. Mass deposition rates and cooling radius are from (1)
Peres et al. (1998) and (2) White et al. (1997). Dust masses are
evaluated from 60$\mu$m data compiled in Edge (2001), assuming
T$_{dust}$=35K.}
\label{results} 
\end{table*} 
\normalsize  
\section{Discussion}   
\subsection{Origin of the cold gas}  Recent
X-ray observations by Chandra, see \cite{Fab02}, \cite{Voi02} and
XMM-Newton (\cite{Pet01}, \cite{Tam01}) have confirmed the presence of
radial gradients in temperature in the cores of several clusters
of galaxies. Even if results from the high spectral resolution
Reflection Grating Spectrometer (RGS) on XMM-Newton do not show
evidence (from Fe XVII) for gas cooling at temperature lower than 1-2
kev, millimetric emission of a cold gas component is detected in the
center of several galaxy clusters.  Added to recent data from
\cite{Edg01}, new detections of molecular gas in cooling flow galaxies
is of great interest. But questions persist on the origin of this cold
phase and its place in a gas infall scenario.  \subsection{Optically
thick cold clouds}  
\begin{figure}[h] \centering \centerline{\epsfig{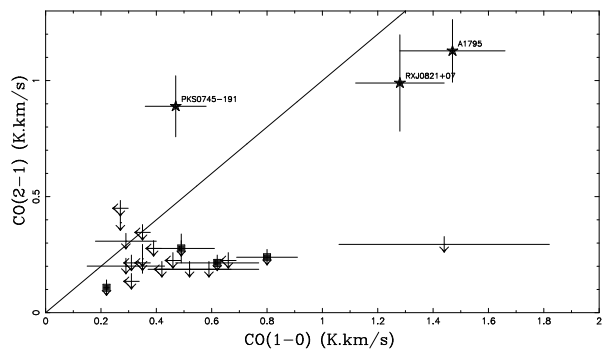}} \caption{CO(2-1)
versus CO(1-0) corrected from the effect of different beam sizes. The
straight line corresponds to line emission
equality.} \label{coco} \end{figure}

In an optically thick medium which is the case
here, the CO(2-1)/CO(1-0) ratio should be about or less than one, if
we assume the same excitation temperature for the two CO energy
levels. On Fig. \ref{coco} is plotted the CO(2-1) versus CO(1-0)
intensity (in K.km.s$^{-1}$) for the galaxies observed here. The
straight line indicates their equality. CO(2-1) intensities have been
multiplied by a beam correcting factor $\sim$4 and by the relative
beam efficiencies (0.52/0.91)/(0.75/0.95)=0.72 to be compared to the
CO(1-0) ones. The preliminary plot indicates that the CO(2-1) line is
in fact lower than the CO(1-0) one; this is in general the case for
sub-thermally excited gas, in nearby galaxies (e.g. \cite{Bra92}). The
CO lines ratio are consistent with an optically thick gas. The medium
considered here is certainly far more complex, probably inhomogeneous
and multi-phase. It might be a mixing of diffuse gas and denser
clumps, and the diffuse medium might be dominating the emission, while
thick and small clouds could enclose a larger quantity of hydrogen
mass than
estimated.    \subsection{Origin of cold
gas}  
\begin{figure}[htbp] 
\centering 
\epsfig{figure=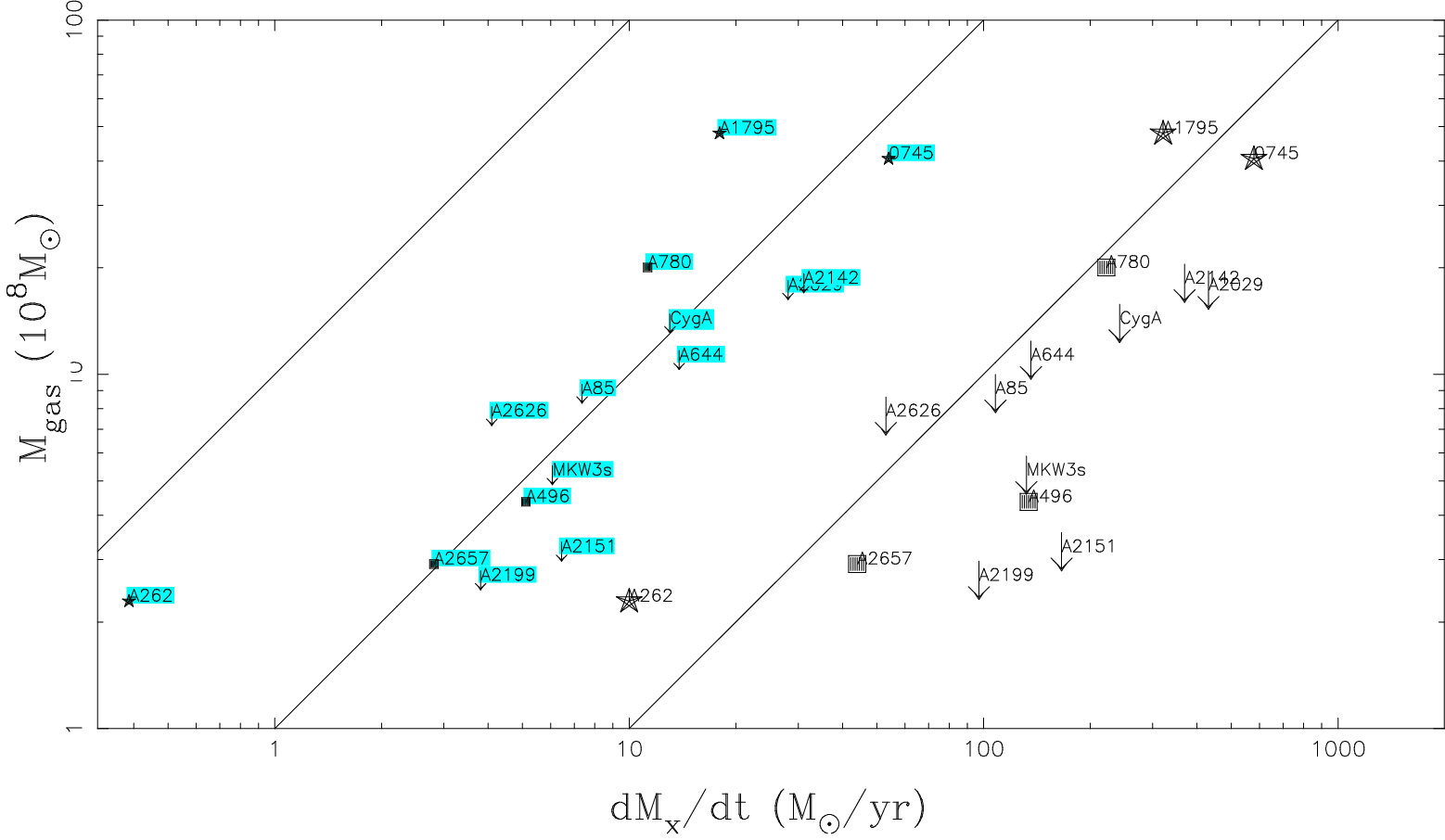,width=4.5cm,angle=-90} 
\caption{M$_{gas}$
from CO observations with respect to $\dot{M}$ the mass deposition
rates deduced from Einstein X-ray data by \cite{whi97}. Straight lines
from right to left are
for M$_{gas}$=1\%$\times\dot{M}\times$1Gyr, M$_{gas}$=10\%$\times\dot{M}\times$1Gyr
and M$_{gas}$=$\dot{M}\times$1Gyr}
\label{mx2not} 
\end{figure}

\begin{figure}[htbp] 
\centering 
\epsfig{figure=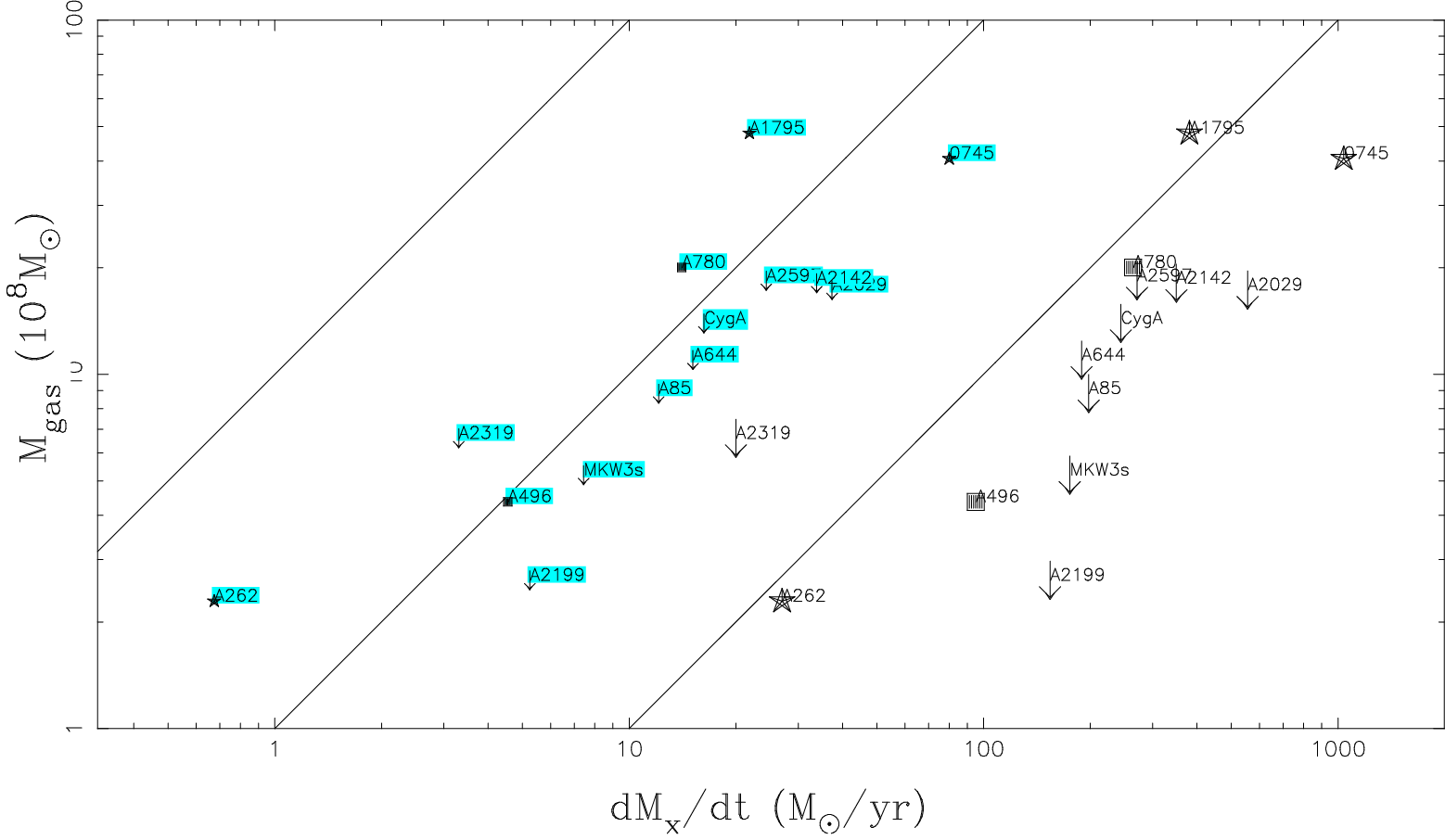,width=4.5cm,angle=-90} 
\caption{M$_{gas}$
from CO observations with respect to $\dot{M}$ the mass deposition
rates deduced from ROSAT X-ray data by
\cite{Per98}.} 
\label{mx2not2} 
\end{figure}

If the molecular gas is formed by the cooling flow, we should
see a correlation between the detected cold gas masses and the
X-ray determined mass deposition rate, as shown by \cite{Edg01}. But
there is a quite large dispersion in the $\dot{M}_X$ values because of
the different methods used in the literature, see
\cite{Grab90}, \cite{Bre90}, \cite{whi97}, \cite{Per98} or
\cite{All00}. To test this, we have compared cold gas masses found
here with mass deposition rates evaluated thanks to an Einstein
Observatory X-ray image deprojection analysis made by \cite{whi97},
see Fig. \ref{mx2not}. The comparison is also done with $\dot{M}$
issue from a ROSAT observatory spatial analysis by \cite{Per98} as
shown on Fig. \ref{mx2not2}. These two samples of mass deposition
rates, evaluated from X-ray data, have been chosen because they
contain the largest number of sources in common with the clusters
observed here in CO. For a correlation trend to be relevant, the aim
was (i) to have a high number of sources observed in both CO and
X-ray, with regards to the faint detection level in CO and (ii) to
compare M(H$_2$) with $\dot{M}$ derived from one method only (with the
same criteria for all sources). We can see a trend of correlation do
appear, even if there are very few data points. This confirms the
relation between the mass of the cold component and the mass
deposition rate already noticed in Edge (2001). Galaxies for which
measurements have been possible lie close to
M$_{gas}$=1\%$\times\dot{M}\times$1Gyr (large symbols). Then, assuming
that simple models of a multiphase flow would lead to an integrated
mass deposition profile of the
form $\dot{M}$($<$r)$\propto$r$^\alpha$, with $\alpha\sim$1. We
have re-evaluated what would be the $\dot{M}$ inside the 30m
telescope radius with a simple scaling by the cooling radius to the CO
radius ratio. The correlation still appear but the cold gas masses
detected are now close to M$_{gas}$=$\dot{M}\times$10\%$\times$1Gyr
(small symbols with gray background). These mass deposition rates
have probably been overestimated by about a factor 5-10, see
\cite{Mcn02}, as suggested by the recent X-ray observations by Chandra
and XMM-Newton (e.g. Abell 1795 in \cite{Fab02}, Abell 2199
in \cite{Joh02} or Abell 496 in \cite{Dup03}). Taking into account
the uncertainty on the conversion factor between H$_2$ and CO,
as discussed above, the correlation is in accordance with a
cooling scenario in which hot gas lead to cold substructures at rate
deduced by X-ray observations and detected here in CO (for an assumed
age of the cooling is a few Gyr in the central
regions).  

But many galaxies in
cooling flow clusters observed here, do not show CO emission
lines. Given the faint emission temperature, it is possible that the
cold gas is present but its radiation is below the detection limit. An
alternative is the gas is not cooled identically in all clusters of
galaxies centers, depending on the environment of the cooling flow
(existence or not of an AGN for example).  
\subsection{Heating by
AGN}  
\begin{figure}[h] 
\centering 
\centerline{\epsfig{figure=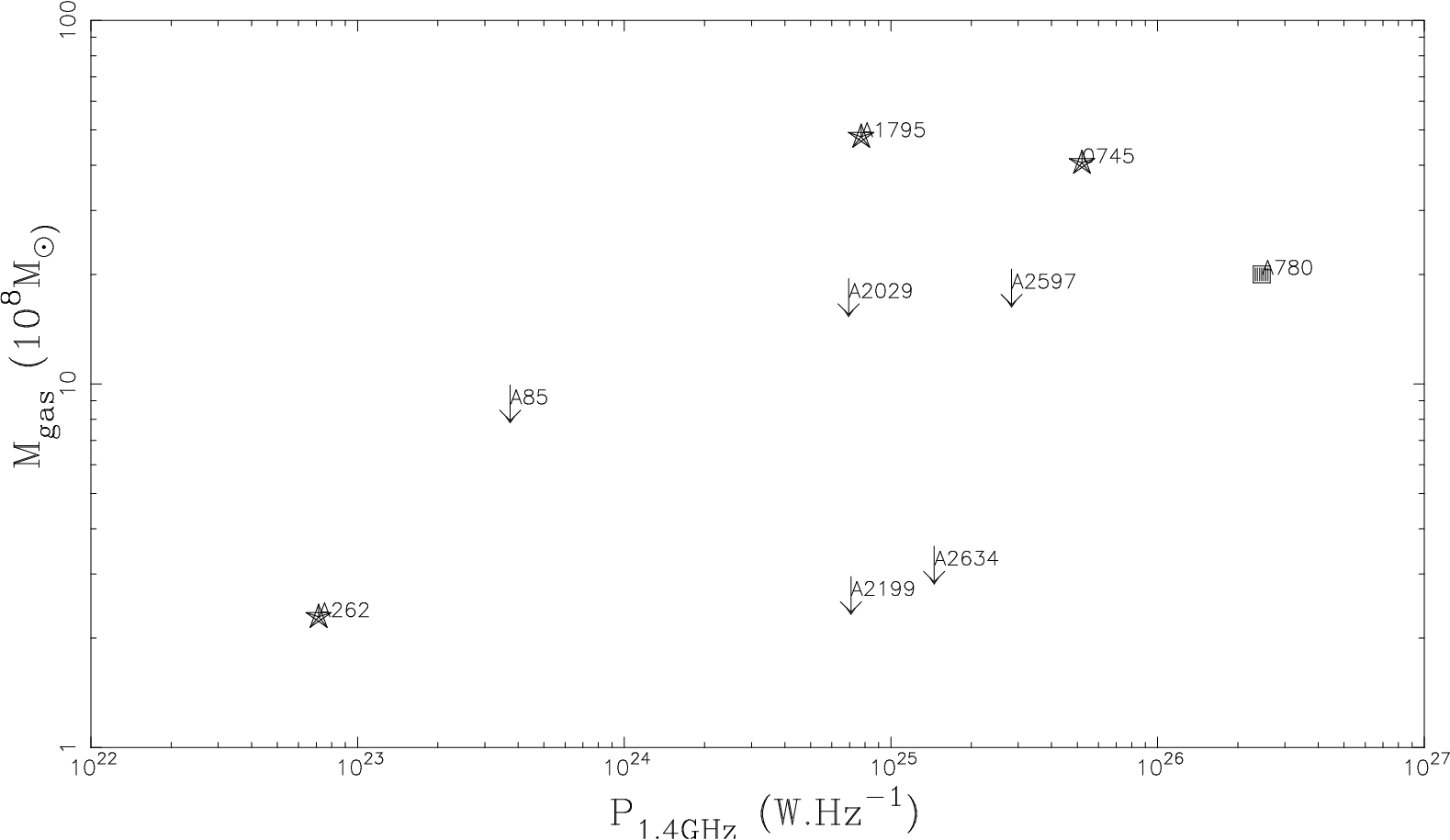,width=4.5cm,angle=-90}} 
\caption{M$_{GAS}$
versus radio power at 1.4 GHz. There is no clear correlation between
molecular gas masses and radio flux.} 
\label{mrnot}
\end{figure}

Many studies developed recently are taking into account
heating mechanisms in cooling flows that could slow down the cooling
and eventually stop it, which might explain the lack of CO emission
in some of the galaxies observed here. Important absorption could
also hide the gas lying in a colder X-ray phase. Physical conditions
in the central regions are certainly very complex, and simple cooling
appears to be insufficient to explain multi-wavelength
observations. Chandra images from \cite{Fab02b} or \cite{Joh02} have
pointed out holes coincident with radio lobes and cold fronts showing
the interaction between the radio source and the intra-cluster medium
(e.g. in Hydra A, the Perseus cluster, Abell 1795, Abell 2199, the
Virgo cluster). It seems that the radio source and jets could heat the
gas with shocks and significantly decrease the cooling rates
(\cite{Dav01}, \cite{Bru02}). Besides, 71 \% of central cD galaxy in
cooling flow clusters show a strong radio activity compared to 23\%
for non-cooling flow cluster cDs (\cite{Bal93}). No correlation has
been found here between molecular gas masses and radio power at 1.4
GHz (see fig. \ref{mrnot}). Nevertheless, it seems that for faint
radio sources, the power at 1.4GHz increases with the cold molecular
gas mass detected, and for stronger radio sources, M$_{gas}$
decreases when the radio power increases. This is consistent with
a self-regulated heating model powered by a central AGN (as suggested
by \cite{Boh01}). But heating the ICM is certainly due to the radio
lobes expansion whose energy is not only linked to the radio
power.  
\subsection{Correlations with H$_{\alpha}$ and M$_{dust}$}  
\begin{figure}[h] \centering \centerline{\epsfig{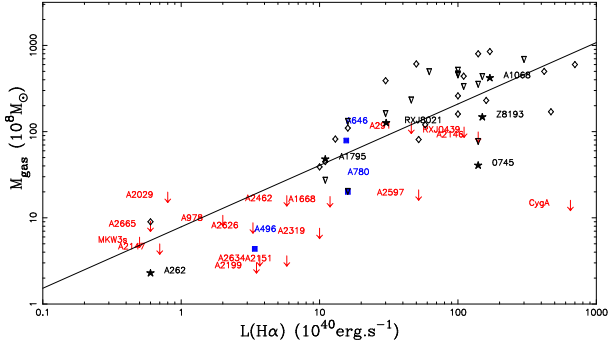}} \caption{M${gas}$
deduced from our CO observations versus H$_{\alpha}$. The straight
line here is a fit to the detection and hint of detection points. Data
from \cite{Edg01} are plotted.  Diamond shaped are the detections and
triangles are the upper limits. Upper limits were not taken into
account in the fit. There is a clear correlation between H$_\alpha$
luminosity and cold gas masses, in agreement with previous
detections.} \label{halpa2not} \end{figure}

The
molecular hydrogen mass appears to be correlated with the amount of
ionized gas (see Fig. \ref{halpa2not}). Excitation processes leading
to the gas emission could be the same for the two
phases. The H$\alpha$ emission could come from shocked cooling
gas. The gas may also have been ionized from massive stars born in a
starburst triggered by the cooling flow. The best linear fit between
the two components is plotted including previous detections of
\cite{Edg01}, (Figure \ref{halpa2not}). Heating by a young star
population is often suggested (see \cite{Joh02}). In that sense,
gradients of metallicity deduced from Chandra observations could be
explained by SN Ia injection of metal in the central galaxy (with the
condition of some exchange of the gas at different radius, and so a
possible mixing of different phases of the gas if they are
present). The cold gas detected here might be a reservoir available
for such a star
formation process.    

\begin{figure}[htbp]
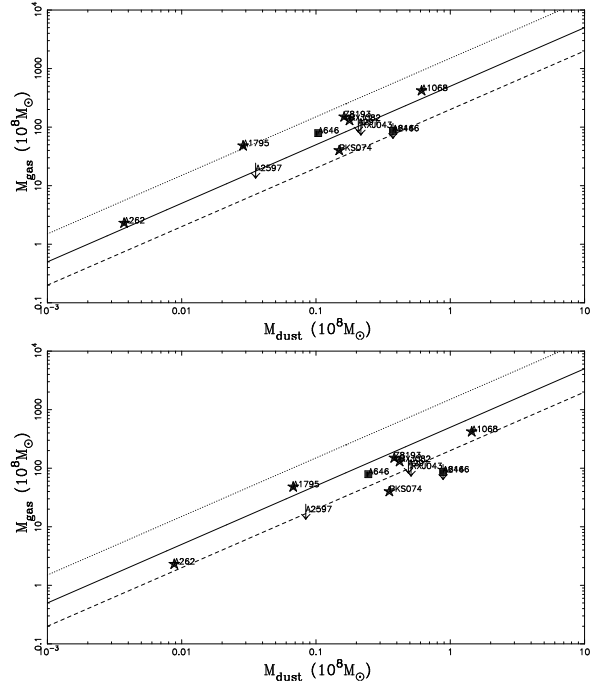
 
\centering 
{\epsfig{figure=./mdust.ps,width=4.5cm,angle=-90}} 
\centering 
{\epsfig{figure=./mdust2.ps,width=4.5cm,angle=-90}}
\caption{M$_{gas}$ molecular hydrogen mass given by CO measurements
as a function of M$_{dust}$ obtained from IRAS data.  The dust
mass highly depends on the dust temperature which is unknown : the
dust temperature is estimated to 35K on the upper figure and to 40K on
the figure below. The straight lines represent gas to dust ratio of
200 (dashed), 500 (full) and 1500 (dotted). The temperature
uncertainties exclude any clear conclusion about the cold gas
origin.} 
\label{rapdustnot} 
\end{figure}

The intracluster gas
should be depleted in dust, at a given metallicity, since in the ICM
environment and its physical conditions, the sputtering time of dust
is much shorter than the dynamical time. The gas coming from a cooling
flow, already at low metallicity, is thought to have a large relative
depletion in dust. Therefore, the expected ratio between the CO
measured gas content and the dust content from its submillimeter or
far-infrared emission is large. Dust masses, derived from IRAS are
evaluated for two assumed dust temperatures T$_{dust}$ = 35 and 40
K. Theses masses are compared to cold gas masses on Fig.
\ref{rapdustnot}. It is important to notice how much dust mass highly
depends on dust temperature. The gas-to-dust ratio for both dust
temperature are high, but below $\sim$2.10$^3$. However, only IRAS
data have been used here, tracing the warm dust. Significant amount of
cold dust might be present as suggested by JCMT SCUBA detections in
Abell 1835, Abell 2390 (Edge et al. 1999). More longer wavelength
observations at 850 $\mu$m tracing this cool dust would be of great
interest. Nevertheless the mass to dust ratio found here are not
incompatible with a cooling flow origin of the molecular gas. Besides,
there is a trend of correlation between cold gas masses and dust
masses, but with a large dispersion. Infrared emission might be tied
to star formation. In that sense, Fig. \ref{rapdustnot} could also be
interpreted as a possible correlation between gas content and star
formation, probably very active as we have seen
previously.   

\subsection{Excited H$_2$
detected in NIR}  Some of our sample galaxies have strong molecular
hydrogen emission in the 2 $\mu$m 1-0 S(1) line,
(\cite{Els92},1994). This excited gas is thought to be associated to
the cooling flow, since it is not detected in non-cooling flow
galaxies of similar-type (\cite{Jaf97}). They reveal dusty nebular
filaments, very similar to those detected in early type galaxies in
small groups e.g.  \cite{Gou98}, and in interacting gas-rich
galaxies. The filaments are extended over kiloparsecs, and their
heating source is not known. Recently, \cite{Don00} have mapped the
kpc-size filaments in vibrationally-excited H$_2$ in the cores of
galaxies centers of Abell 2597 and PKS 0745-191 with high spatial
resolution. They have also found dust lanes which are optically thick
to 1.6$\mu$m emission, confined to the central few kpcs. Excited H$_2$
produced directly by the cooling flow seems difficult, since H$_2$ is
much too luminous, by at least 2 orders of magnitude. It cannot be AGN
photoionization or fast shocks because the H$\alpha$/H$_2$ ratios are
too low. Extremely slow shocks ($<40$ km s$^{-1}$) produce
significantly higher H$_2$/H$\alpha$ ratios than do fast shocks, and
are more consistent with the observations. But slow shocks are less
efficient. The most likely solution is UV irradiation by very hot
stars, implied by a star formation rate of only a few solar masses per
year. A recent survey of H-band and K-band spectra in 32 central
cluster galaxies have led to 23 detections in rovibrationnal H$_2$
lines, see \cite{Edg02}, and UV fluorescence excitation is ruled
out. The molecular hydrogen is more probably thermally excited in
dense gas with density exceeding 10$^5$ cm$^{-3}$ at temperature
evaluated between 1000-2500 K. Young stars heating a population of
dense clouds is invoked (\cite{Wil02}). According to the authors,
these dense regions might be self gravitating clouds deposited
directly by the cooling flow or confined in high pressure behind
strong shocks.  Correlation are also shown between H$_\alpha$ emission
lines, warm H$_2$ rovibrationnal lines and cold millimetric emission
lines suggesting related exciting mechanisms of these different phases
of the gas. The large masses of excited H$_2$, around
10$^{5-6}$M$_\odot$ could suggest that the cold molecular gas mass
could have been underestimated (because of a lower metallicity for
example) or is hidden in optically thick dense clouds, see
\cite{Fer02} for a discussion of the physical conditions within dense
cold clouds in cooling flows.\\ \\ How much gas is deposited
in cooling flows is still an open question. The gas cooling in the
flow is probably multi-phase, and there are hints the CO detected here
is the residual of the cooled gas. But this cold gas emission could
also be due to subcluster structures, gas stripped from
neighbouring galaxies or galactic clouds not seen until now and heated
by mechanisms linked to the flow, like shocks or
starburst. More investigations are required to explore the properties
of this important component in cooling flow cluster cores. The study
of the morphological structure of the cold gas and especially its
dynamics will help to confirm its place in the flow. High resolution
maps, obtained thanks to the IRAM millimeter interferometer, have
been obtained for Abell 1795 in CO(1-0) and CO(2-1). These maps show
an extended emission of the cold gas (Salom\'e \& Combes, in
prep). They underline the possible link between the cold gas detected
with the 30m telescope and the cooling gas seen at higher
energy. Recent OVRO observations by \cite{Edg03} also show CO(1-0)
emission maps in 5 cooling flow clusters of galaxies : A1068,
RXJ0821+07, Zw3146, A1835 and RXJ0338+09. The authors conclude the gas
previously detected with the single dish telescope is confined in the
central region. More Plateau de Bure interferometric observations with
higher sensitivity and spatial resolution are in progress now in
RXJ0821+07 to see whether the cold gas is extended (as for Abell 1795)
or centrally concentrated around the cD (as suggest the
OVRO observations). Interferometric observations on a wider sample of
CO detected cooling flow have now to be lead in order to explore
the similarities and differences between clusters and definitively
confirm the detection of the cold residual in cooling
flows.  \section{Conclusions}  A sample of 32 cooling flow clusters of
galaxies, selected on their mass deposition rate, and their H$\alpha$
luminosity, have been observed in both CO(1-0) and CO(2-1) emission
lines. In total 6 clear detections are claimed, with 4 other possible
detections.  Molecular hydrogen mass estimates have been deduced for
these galaxies and upper limits have been computed for the other ones.
The derived M(H$_2$) are up to 10$^{10}$M$_\odot$ in the 22 central
arcseconds observed with the 30m telescope (that is typically the
central $\sim$23kpc region at z=0.05). These masses appear to be
related to the cooling rate deduced from X-ray data : there is a trend
of correlation with $\dot{M_X}$ results, and no longer large
discrepancies between the mass deposition rates and the cold gas
masses (according to recent mass deposition rates reevaluation from
Chandra and XMM-Newton). The apparent gas-to-dust ratio, derived from
the CO emission and dust far-infrared emission is larger for the gas
in cooling flow galaxies than in normal spirals, but uncertainties
about the dust temperature preculdes any clear conclusions. The best
correlation is between the cold gas masses and the H$\alpha$
luminosities, which confirms the result of Edge (2001). Further work
is to be done now to confirm that CO lines, revealed by single dish
millimetric observations, are tracers of the long searched cold phase
in cooling flows. In this context, more interferometric observations
in CO(1-0) and CO(2-1) are required.  \begin{acknowledgements} It is a
pleasure to thank the IRAM-30m staff for their support during 
observations and data reduction, especially with the new 4MHz 
filter-bank. We also thank Alastair Edge for his constructive 
refereing. \end{acknowledgements}  
\begin{figure*} 
\caption{CO(1-0)
and CO(2-1) emission lines observed with the IRAM 30m telescope. On the Y-axis, main beam
temperature (in mK)  versus velocity (in km/s) on the X-axis .} 
\begin{center} 
\includegraphics[scale=0.95]{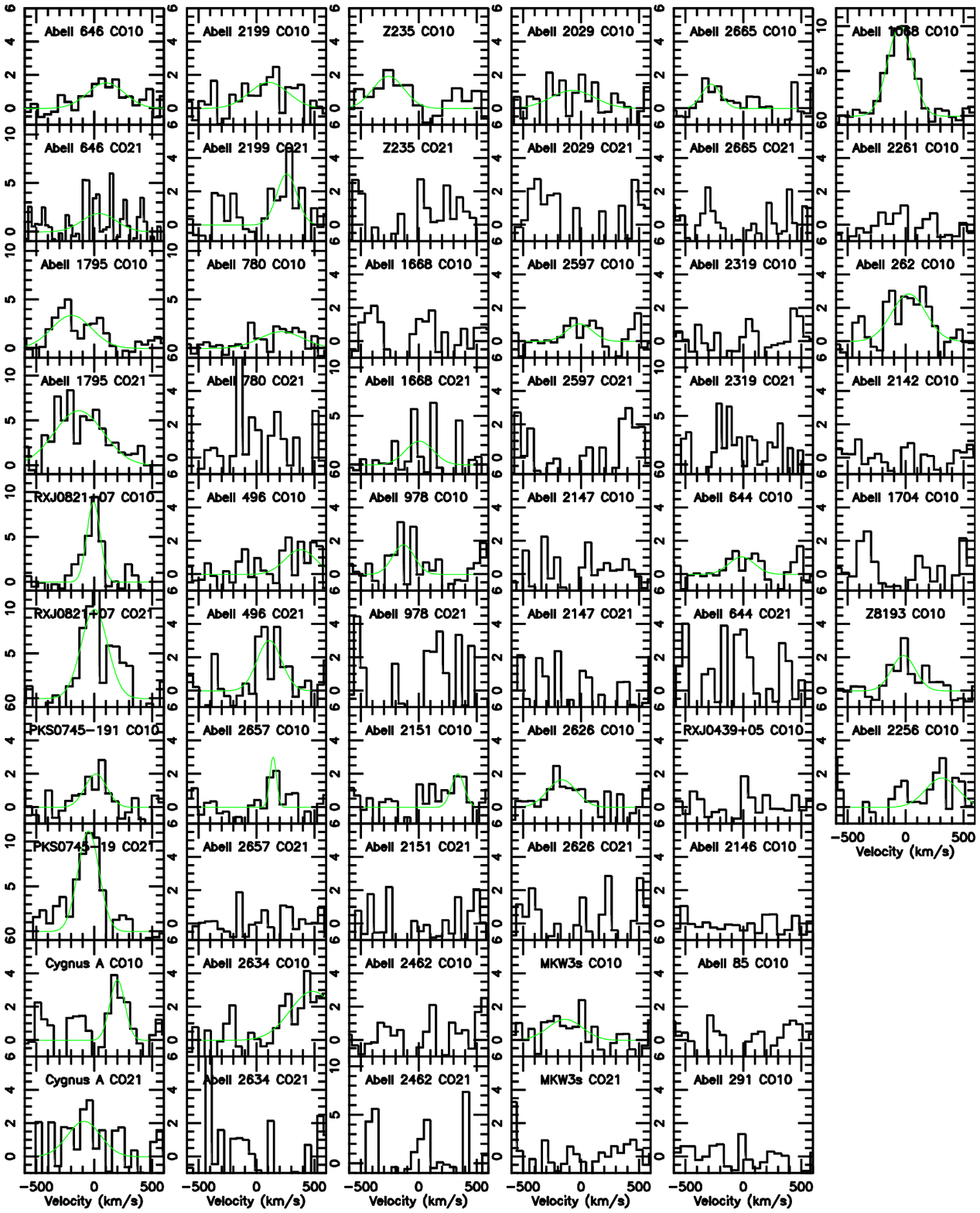} 
\end{center} 
\label{spectres} 
\end{figure*}  

\begin{thebibliography}{} 
\bibitem[Antonucci
et al.  (1994)]{Ant94}Antonucci R., Barvainis R.: 1994, AJ 107,
448 \bibitem[Allen (2000)]{All00} Allen S. W.: 2000, MNRAS 315,
269 \bibitem[Baum \& O'Dea (1991)]{Bau91} Baum, S. A.; O'Dea, C. P. :
1991 MNRAS, 250, 737B \bibitem[Bayer-Kim et al. (2002)]{Bay02}
Bayer-Kim C. M., Crawford C. S., Allen S. W., Edge A. C. and Fabian
A.C. : 2002 MNRAS 337, 938 \bibitem[Bonamente et al.  2001]{Bon01}
Bonamente, M., Lieu, R., Mittaz, J. P. D.: 2001 ApJ 546, 805 \& 547,
L7 \bibitem[B\"{o}hringer et al.  (2001)]{Boh01} B\"{o}hringer
H., Matsushita K., Ikebe Y. : astro-ph/0111113 \bibitem[Braine \&
Combes 1992]{Bra92} Braine J., Combes F.: 1992, A\&A 264,
433 \bibitem[Braine \& Dupraz (1994)]{Bra94}Braine J., Dupraz C.:
1994, A\&A 283, 407 \bibitem[Bregman et al.  (1990)]{Bre90} Bregman
J., Mc Namara B.R., and O'Connell W.: 1990, ApJ 351,
406 \bibitem[Brighenti \& Mathews 2002]{Bri02} Brighenti F. \&
Mathews W.G. : 2002, astro-ph/0203409 \bibitem[Br\"uggen et
al. (2002)]{Bru02} Br\"uggen M., Kaiser C. R., Churazov, E., Ensslin,
T.  A.: 2002 MNRAS, 331, 545B \bibitem[Burns et al.  1981]{Bur81}Burns
J.O., White R.A., Haynes M.P.: 1981, AJ 86, 1120 \bibitem[Ball et al.
1993]{Bal93}Ball R., Burns J.O., Loken C., : 1993, AJ
105,53 \bibitem[Cowie et al.  (1983)]{Cow83} Cowie, L. L., Hu,
E. M., Jenkins, E. B., York, D G.  : 1983 ApJ, 272,
29C \bibitem[Crawford et al.  (1999)]{Cra99} Crawford C.S., Allen
S.W., Ebeling H., Edge A.C, Fabian A.C.: 1999 MNRAS, 306,
857 \bibitem[David et al.  (1993)]{Dav93} David, L. P., Slyz, A.,
Forman, W and Vrtilek, S.  D.: 1993, ApJ 412, 479 \bibitem[David et
al.  2001]{Dav01} David, L. P., Nulsen, P. E. J., McNamara, B. R. et
al.: 2001, ApJ 557, 546 \bibitem[Donahue et al.  (2000)]{Don00}Donahue
M., Mack J., Voit G.M, Sparks W., Elston R., Maloney P.R.: 2000, ApJ
545, 670 \bibitem[Dunne \& Eales (2001)]{Dun01} Dunne L. and Eales
S.A. : 2001, MNRAS 327, 697 \bibitem[Dupke \& White (2003)]{Dup03}
Dupke R. and White III R.E.: 2003, ApJ 583, L13 \bibitem[Durret et al.
1994]{Dur94}Durret F., Gerbal D., Lachi\`eze-Rey M., Lima-Neto G.,
Sadat R.: 1994, A\&A 287, 733 \bibitem[Dwarakanath et
al. 1995]{Dwa95}Dwarakanath K.S., Owen F.N., van Gorkom J.H.: 1995,
ApJ 442, L1 \bibitem[Edge (1999)]{Edg99} Edge A.C., Ivison R.J., Smail
Ian, Blain A.W., \& Kneib, J.P.: 1999, MNRAS 306, 599 \bibitem[Edge
(2001)]{Edg01} Edge A.C.: 2001, MNRAS 328, 762 \bibitem[Edge et
al. (2002)]{Edg02} Edge A.C., Wilman R.J., Johnstone R.M., Crawford
C.S., Fabian A.C., Allen S.W.  : 2002, astro-ph/0206379 \bibitem[Edge
\& Frayer (2003)]{Edg03} Edge A.C., Frayer, D.T. : 2003, ApJ, 594,
L13 \bibitem[Elston \& Maloney 1992]{Els92} Elston, R. \&
Maloney, P. 1992, BAAS, 181, 118.11 \bibitem[]{Els94} Elston, R. \&
Maloney, P. 1994, in ``Infrared Astronomy with Arrays: The Next
Generation'', ed. I. S. McLean (Kluwer: Dordrecht),
p. 169 \bibitem[Ettori et al. (2002)]{Ett02} Ettori, S., Fabian,
A. C., Allen, S. W., Johnstone, R. M : 2002 MNRAS 331,
635E \bibitem[Fabian 1994]{Fab94} Fabian A.C.: 1994, ARAA 32,
277 \bibitem[Fabian et al. 2001]{Fab01} Fabian A.C., Sanders J.S.,
Ettori S. et al: 2001, MNRAS 321, L33 \bibitem[Fabian 2002]{Fab02}
Fabian A.C. : 2002, astro-ph/0201386 \bibitem[Fabian et
al. (2002)]{Fab02b} Fabian A.C., Celotti A., Blundell K.M., Kassim,
N.E. and Perley R.A., MNRAS 331, 369 \bibitem[Falcke et
al. 1998]{Fal98} Falcke, H., Rieke, M. J., Rieke, G. H., Simpson, C.,
Wilson, A. S., 1998, ApJ 494, L155 
\bibitem[Ferland
et al. 1994]{Fer94} Ferland G.J., Fabian A.C., Johnstone R.M.: 1994,
MNRAS 266, 399 \bibitem[Ferland et al. (2002)]{Fer02} Ferland G.J.,
Fabian A.C., Johnstone R.M.: 2002, astr-ph/0203052 \bibitem[Grabelsky
\& Ulmer (1990)]{Grab90} Grabelsky D.A., \& Ulmer M.P.: 1990, ApJ 355,
401 \bibitem[Goudfrooij \& Trinchieri 1998]{Gou98}Goudfrooij,
P. \& Trinchieri, G. : 1998, A\&A, 330, 123 
\bibitem[Jaffe \& Bremer 1997]{Jaf97}Jaffe W., Bremer M.N. 1997:
MNRAS 284, L1 \bibitem[Johnstone et al. (2002)]{Joh02} Johnstone R.M.,
Allen S.W., Fabian A.C. and Sanders J.S. : 2002,
astro-ph/020207 \bibitem[Lazareff et al. (1989)]{Laz89}Lazareff B.,
Castets A., Kim D.W., Jura M.: 1989, ApJ 335, L13 \bibitem[Lieu et
al. (1996,1999)]{Lie96}Lieu R., Mittaz J.P.D., Bowyer S., et al.~
1996, ApJ 458, L5 \bibitem[]{Lie99} Lieu R., Bonamente, M., Mittaz,
J. P. D.: 1999, ApJ 517, L91 
\bibitem[McNamara et al. 1990]{Mcn90}McNamara B.R., Bregman
J.N., O'Connell R.W.: 1990, ApJ 360, 20 \bibitem[McNamara et
al. (1994)]{Mcn94}McNamara B.R., Jaffe W.: 1994, A\&A 281,
673 \bibitem[McNamara et al. (2002)]{Mcn02}McNamara B.R.,
2002, astro-ph/0202199 \bibitem[Mittaz et al. (1998)]{Mit98}Mittaz
J.P.D, Lieu R., Lockman F.J.: 1998, ApJ 498, L17 
\bibitem[O'Dea et al. (1994)]{Ode94}O'Dea C.P., Baum
S.A., Gallimore, J.F. 1994b, ApJ, 436, 669 
\bibitem[O'Dea et
al. (1994b)]{Ode94b}O'Dea C.P., Baum S.A., Maloney P.R., Tacconi
L.J. and Sparks W.B., ApJ 422, 467 \bibitem[Owen et al. (1995)]{Owe95}
Owen F., Ledlow M.J., Keel W.C.: 1995, AJ 109, 14 \bibitem[Peres et
al. (1998)]{Per98} Peres C.B., Fabian A.C., Edge A.C., Allen S.W.,
Johnstone R.M. and White D.A.: 1998, MNRAS 298, 416 \bibitem[Peterson
et al. 2001]{Pet01} Peterson J.R., Paerels F.B.S., Kaastra J.S.,
Arnaud M., Reiprich T.H., Fabian A.C., Mushotzky R.F., Jernignan
J.G. and Sakelliou I. : 2001, A\&A 365, L104 \bibitem[Pfenniger \&
Combes 1994]{Pfe94}Pfenniger D., Combes F., 1994, A\&A, 285,
94 \bibitem[Shostak et al. 1983]{Sho83}Shostak G.S., van Gorkom
J.H., Ekers R.D. et al.: 1983, A\&A 119, L3 \bibitem[Tamura et
al. 2001]{Tam01} Tamura T., Kaastra J.S., Peterson J.R., Paerels
F.B.S., Mittaz J.P.D., Truddolyubov S.P., Stewart G., Fabian A.C.,
Mushotzky R.F., Lumb D.H., Ikebe Y. : 2001, A\&A 365,
L87 \bibitem[Tucker et al. 1997]{Tuc97}Tucker W., David, L. P.: 1997,
ApJ 484, 602 \bibitem[Valentijn \& Giovanelli 1982]{Val82}Valentijn
E.A., Giovanelli R.: 1982, A\&A 114, 208 
\bibitem[Voigt et al. (2002)]{Voi02} Voigt L.M., Schmidt R.W.,
Fabian A.C., Allen S.W.  and Johnstone R.M. : 2002,
astro-ph/0203312 
\bibitem[White et
al. (1997)]{whi97} White D.A., Jones C. and Forman W.: 1997, MNRAS
292, 419 \bibitem[Wilman et al. 2002]{Wil02}Wilman R.J., Edge A.C.,
Johnstone R.M., Fabian A.C., Allen S.W.  and Crawford C.S. :
astro-ph/0206382 \bibitem[Wu \& Hammer 1993]{Wu93}Wu X.P., Hammer F.:
1993, MNRAS 262, 187 
\end{thebibliography}
\end{document}